\documentclass[11pt,twoside
]{article}

\usepackage{baaa2008}
\usepackage{graphicx}
\usepackage{subfigure}
\usepackage{psfrag}
\usepackage{amssymb}
\usepackage[spanish,activeacute,english]{babel}
\usepackage[latin1]{inputenc}
\usepackage[T1]{fontenc} 
\usepackage{ae,aecompl} 
\usepackage{latexsym}
\usepackage{verbatim}
\usepackage{amsmath}
\usepackage{stmaryrd}
\usepackage{amsfonts}
\usepackage{amssymb}
\usepackage{wasysym}
\usepackage[colorlinks=true,dvips]{hyperref}

\begin{document}
\myselectspanish
\vskip 1.0cm
\markboth{ A. Collado, R. Gamen \& R. Barb\'a}%
{Estrellas WN en la V\'ia L\'actea: Campa\~na 2007-2008}

\pagestyle{myheadings}
\vspace*{0.5cm}
\parindent 0pt{PRESENTACION MURAL}
\vskip 0.3cm
\title{Estrellas Wolf-Rayet de tipo WN en la V\'ia L\'actea: Campa\~na 2007-2008}


\author{A. Collado$^{1}$,
R. Gamen $^{1}$, 
R. Barb\'a$^{1,2}$}

\affil{%
  (1) Complejo Astron\'omico El Leoncito, San Juan, Argentina\\
  (2) Departamento de F\'isica, Universidad de La Serena, Chile\\
}

\begin{abstract} 
We are carrying out a spectroscopic monitoring of Galactic Wolf-Rayet stars,
 in order to detect binary systems. The sample consists of approximately 50
 stars of the Nitrogen sequence (WN) and fainter than $V$=13. The
 observations are made from the 4-m telescope at CTIO, Chile. 
In the following, we present the first results of the 2007-2008 campaign.
\end{abstract}

\begin{resumen}
Estamos llevando a cabo un monitoreo espectrosc\'opico de estrellas 
Wolf-Rayet de nuestra galaxia, con el objeto de detectar sistemas binarios. 
La muestra consiste de unas 50 estrellas de la secuencia del Nitr\'ogeno (WN) 
y m\'as d\'ebiles que $V$=13. Las observaciones son realizadas desde el 
telescopio de 4-m de CTIO, Chile. A continuaci\'on, presentamos los primeros
resultados de la campa\~na 2007-2008.
\end{resumen}

\section{Motivaci\'on}
Las estrellas Wolf-Rayet (WR) de Poblaci\'on I, 
tienen vientos estelares muy intensos, densos, y calientes,
que le dan la forma caracter\'istica
a sus espectros, i.e. l\'ineas de emisi\'on intensas y anchas.
Seg\'un la presencia de distintas l\'ineas en el espectro\'optico, 
se clasifican en WN (Nitr\'ogeno), WC (Carbono), y WO (Ox\'igeno).
Los par\'ametros
intr\'insecos de las estrellas WR son muy dif\'iciles de determinar dado 
que muchas de sus propiedades están enmascaradas por sus 
poderosos vientos.

Las estrellas WR son consideradas des\-cen\-dien\-tes de las estrellas
tipo O, aunque
la forma en que se produce esta transici\'on es a\'un motivo de debate.
Hay un subconjunto de las estrellas WN, designadas como WNH,
que se las considera estrellas O muy masivas con fuertes
vientos estelares en una etapa evolutiva a\'un temprana, distintas de las WR
``cl\'asicas'' que queman He en sus n\'ucleos (Conti \& Smith, 2008).

La masa estelar es un par\'ametro astrof\'isico fundamental cuyo
conocimiento, junto al \'indice de p\'erdida de masa, la composici\'on
qu\'imica, y la rotaci\'on,
permite determinar las propiedades y evoluci\'on de las estrellas, en especial
las m\'as masivas (Meynet \& Maeder, 2005). 
El m\'etodo m\'as directo para medir masas estelares es a trav\'es de la
aplicaci\'on de la tercera ley de Kepler.  
Las masas t\'ipicas determinadas en estrellas WR tienen valores entre 9--16\,M$_\odot$
para las WC y 10--83\,M$_\odot$ para las es\-tre\-llas WN
(cf. Crowther 2007).
Las candidatas a ser las estrellas m\'as masivas de la V\'ia L\'actea parecen ser del tipo
WNH, e.g.
las 2 componentes WN6 de NGC3603-A1, con masas absolutas de 116 y 
89\,$M_\odot$ (Schnurr et al. 2008),
y la componente WN6 de WR~21a que result\'o tener una masa m\'inima de 
87 M$_\odot$ (Niemela et al. 2008).
Figer (2005) determina por an\'alisis estad\'istico que se deber\'ia esperar
descubrir estrellas con hasta 150 M$_\odot$, pero los objetos descubiertos al
momento a\'un no llegan a ese l\'imite, por lo que se espera poder hallarlos,
y confirmar (o no) entonces dicho valor.

La fracci\'on de sistemas binarios detectados entre las estrellas WR de
nuestra galaxia es de $\sim40$\% (van der Hucht, 2001),
sin embargo hay al menos dos razones para sospechar que \'este es un
l\'imite inferior.
Una raz\'on es dada por Langer \& Heger (1999), quienes argumentan que la
mayor\'ia de las Supernovas tipo Ib/c ocurren en binarias WR+OB
interactuantes, y dada la estad\'istica derivada de las SN
observadas se esperar\'ia que existan m\'as estrellas WR en sistemas binarios
que simples.
Otra raz\'on es aportada por la observaci\'on de estrellas del tipo WC.
M\'as del 80 \% de las estrellas WC9 y WC8 parecen
pertenecer a sistemas binarios (cf. Tuthill et al. 1999; Monnier et al. 1999).
La pregunta obvia es si esta frecuencia alta de binarias es
intr\'inseca de este sub-tipo o puede ser aplicado a todas las estrellas WR.
Esto, entonces, podr\'ia implicar que la poblaci\'on de estrellas WR est\'a
asociada a binaridad, a trav\'es de la transferencia de masa durante una
determinada fase de un sistema O+O, 
y que las estrellas WR son una clave en la evoluci\'on de estrellas masivas.


Dadas estas consideraciones, 
estamos llevando a cabo un monitoreo espectrosc\'opico de estrellas
WR de nuestra galaxia, con el objeto de detectar variaciones de velocidad
radial que indiquen la presencia de sistemas binarios.
La muestra consiste de unas 50 estrellas de la secuencia del Nitr\'ogeno (WN)
y m\'as d\'ebiles que $V$=13. 
A continuaci\'on, presentamos los primeros resultados de la campa\~na 2007-2008.

\section{Observaciones}

Las observaciones fueron realizadas con el telescopio V. Blanco
de 4-m del Observatorio Inter-Americano de Cerro Tololo (CTIO), Chile,
durante marzo-abril 2007 y abril 2008.
Utilizamos el espectr\'ografo R-C con la red KPGL1, configuraci\'on que
provee un rango espectral de 3650-6700 \AA\ y una dispersi\'on
rec\'iproca de $\sim$ 1 \AA~pix$^{-1}$.

Durante 11 noches de observaci\'on obtuvimos 237 espectros de 44 estrellas WN. 
Las im\'agenes fueron procesadas con las tareas del paquete LONGSLIT de 
IRAF. 
Las velocidades radiales (VR) de las l\'ineas espectrales fueron determinadas
mediante ajustes de funciones gaussianas a los perfiles, con la tarea SPLOT.

\section{Resultados preliminares: Dos nuevas binarias WN+O}
\subsection{WR~62a}

Una de las m\'as conspicuas variables en VR en la muestra observada 
es WR~62a. 
Esta estrella fue descubierta y clasificada como WN5 por Shara et al. (1999),
y no observaron l\'ineas de absorci\'on, sin embargo
en nuestros espectros identificamos l\'ineas en absorci\'on 
H$\alpha$, H$\beta$, H$\gamma$, H$\delta$, He {\sc i} $\lambda$ 4471, y 5875,
y He{\sc ii} $\lambda$ 4200, 4541, y 5411. 

Teniendo en cuenta las variaciones de VR noche a noche, determinamos la periodicidad (mediante un algoritmo similar al descripto pot Bertiau \& Grobben, 1969) de la l\'inea de emisi\'on de He {\sc ii} 4686 \AA\, y obtuvimos un valor del orden de 10 d\'ias. Un resultado similar se obtuvo a partir de otras l\'ineas de emisi\'on. Utilizando este per\'iodo pudimos comprobar que las VRs de las l\'ineas de absorci\'on muestran un movimiento en antifase con las l\'ineas
de emisi\'on, lo que sugiere que se trata de un sistema binario de tipo WN con
una compa\~nera O (siendo el sub-tipo espectral a\'un no determinado). 
Las masas m\'inimas derivadas para las componentes del sistema 
son $\sim$ 40\,M$_\odot$ para la estrella O y $\sim$ 15\,M$_\odot$ para la WN.

La figura~\ref{vr} (izquierda) muestra la variaci\'on de las VRs 
de He{\sc ii} 4686~\AA\, en emisi\'on y el promedio de las
absorciones de He {\sc i} $\lambda$ 4471 y 5875 en funci\'on de
la fase orbital. Las curvas continuas representan la soluci\'on orbital preliminar obtenida.

\subsection{WR~68a}

Otra de las estrellas que presenta grandes variaciones de VR es WR~68a. 
La cual tambi\'en fue descubierta y clasificada por Shara et al. (1999). 
Ellos propusieron un tipo WN7, es decir que tampoco notaron absorciones en este
espectro. En nuestros espectros,
identificamos las l\'ineas de absorci\'on de H$\delta$, H$\gamma$ y H$\beta$, 
He {\sc i} $\lambda$ 4471 y 5875, y He {\sc ii} 4541, las que una vez determinado
el per\'iodo, identificamos como pertenecientes a la compa\~nera, indicando
un tipo espectral O.

La variaci\'on de las VRs de las l\'ineas en el espectro de WR~68a confirman que se
trata de un sistema binario de corto per\'iodo (P del orden de 5 d\'ias) en una
\'orbita con excentricidad no despreciable. Utilizamos el mismo algoritmo para
determinar la soluci\'on orbital considerando las l\'ineas de He {\sc ii}
$\lambda$ 4686 en emisi\'on y el promedio de H$\beta$ y H$\gamma$
en absorci\'on como representando el movimiento de las estrellas WN y 
O. Para este sistema, obtuvimos masas m\'inimas de $\sim$ 38\,M$_\odot$ para la 
componente O y $\sim$ 17\,M$_\odot$ para la WN.

\begin{figure}[!t]
  \centering
  \includegraphics[width=.5\textwidth]{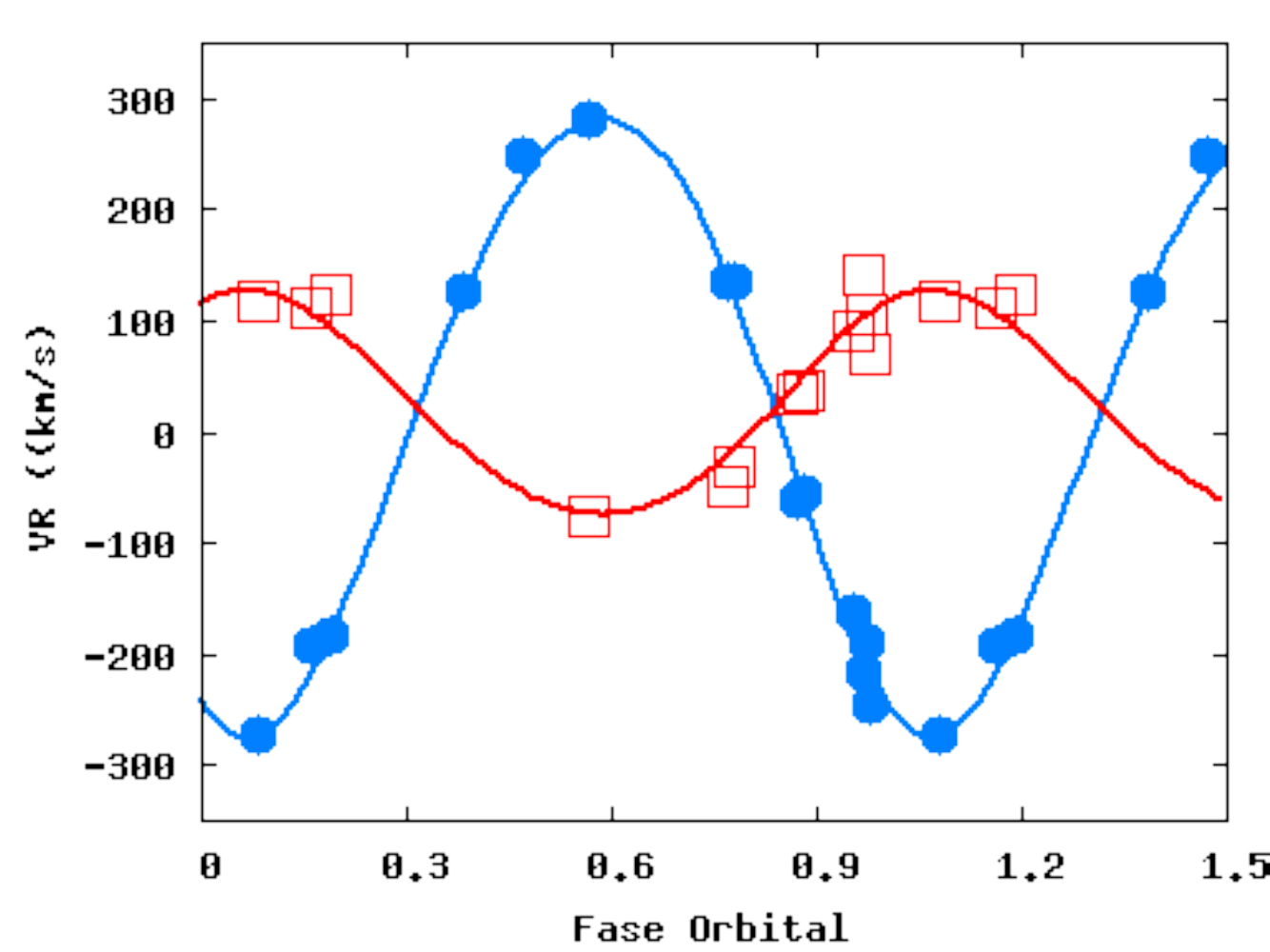}~\hfill
  \includegraphics[width=.5\textwidth]{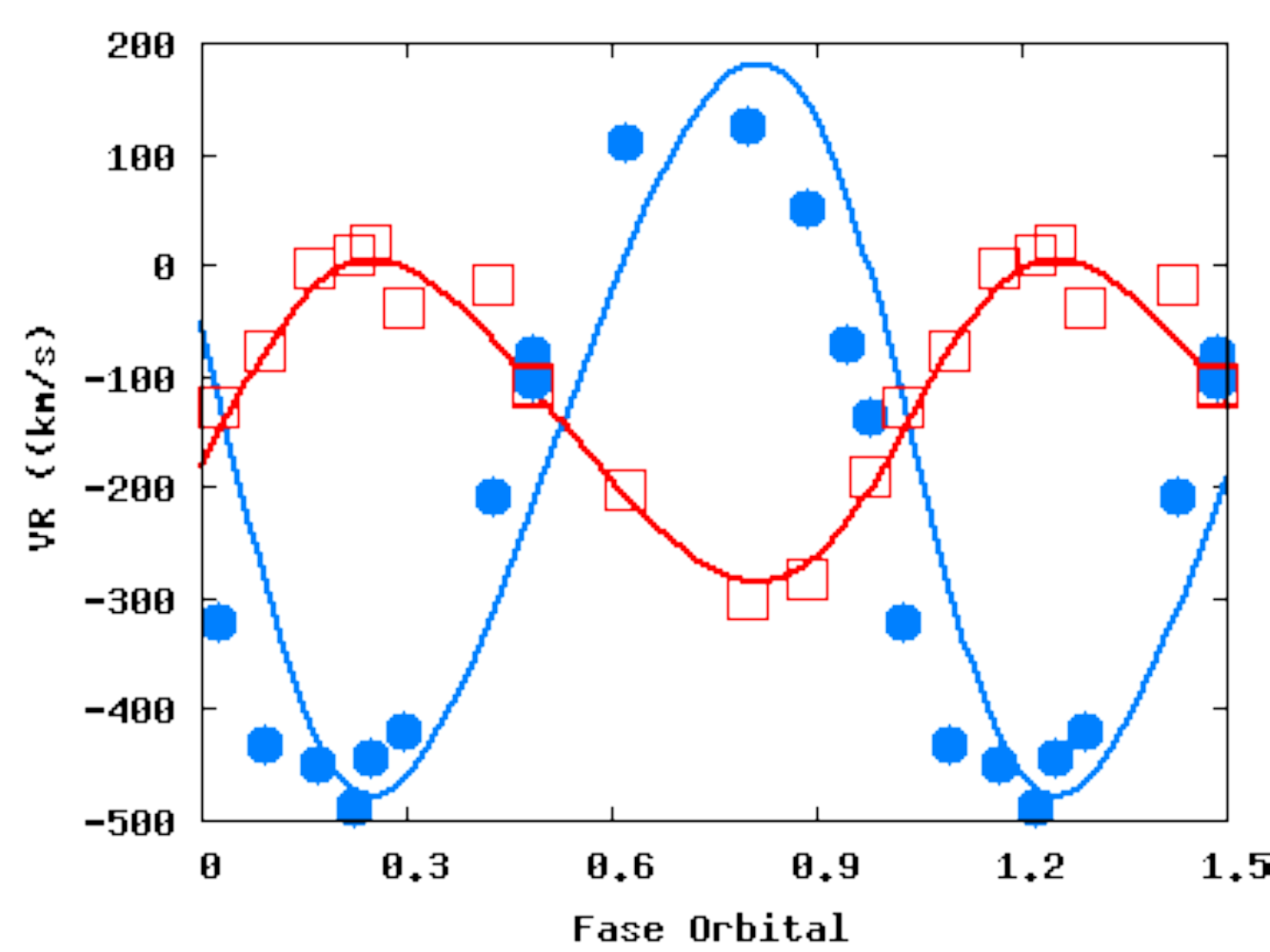}
  \caption{{\it Izquierda:} Variaci\'on de las velocidades radiales de la
l\'inea de emisi\'on de He {\sc ii} (c\'irculos) y el promedio de las absorciones 
(cuadrados) 
como representando el movimiento orbital de las componentes WN y O 
respectivamente en el sistema binario WR~62a.
        \protect\\{\it Derecha:} 
  Idem, para WR~68a. Notar el desfasaje que parece existir en las VRs de la
emisi\'on de He {\sc ii}.}
  \label{vr}
\end{figure}

En la figura~\ref{vr} (derecha) ilustramos las variaciones de VR de las l\'ineas 
utilizadas y, en l\'inea continua, la \'orbita obtenida. En dicha figura, se nota 
un desfasaje en las VRs de la emisi\'on de
He {\sc ii} y las absorciones de H. Este desfasaje no se aprecia si consideramos
la emisi\'on de N {\sc iv}, lo que sugiere una distorsi\'on de los perfiles de las 
l\'ineas de He {\sc ii} provocados por la regi\'on de interacci\'on de vientos.
Este fen\'omeno es observado en otros sistemas WN+O de corto per\'iodo, 
e.g. WR~21 y WR~31 (Gamen 2004). 

\section{Conclusiones}

Descubrimos 2 nuevos sistemas binarios WN+O de corto per\'iodo, WR~62a y WR~68a.
En ambos sistemas pudimos identificar l\'ineas de absorci\'on pertenecientes a
una compa\~nera de tipo espectral O, lo que nos permiti\'o estimar masas m\'inimas 
preliminares
de 40+15\,M$_\odot$, para las componentes O y WN respectivamente en WR~62a y 
38+17\,M$_\odot$ en WR~68a.
Dadas las masas m\'inimas obtenidas y el corto per\'iodo de ambos sistemas, podr\'ian
esperarse variaciones fotom\'etricas debidas a eclipses.
Son necesarias m\'as observaciones espectrosc\'opicas 
para obtener un per\'iodo m\'as preciso en ambos sistemas.

\agradecimientos
Agradecemos al director y staff de CTIO, Chile, por facilitarnos
el uso de sus instalaciones y su hospitalidad.
RG agradece financiaci\'on de FUPACA.
IRAF is distributed by the National Optical
Astronomy Observatories, which are operated by the Association of Universities
for Research in Astronomy, Inc., under cooperative agreement with the National
Science Foundation.

\begin{referencias}

\reference Bertiau F., Grobben J., 1969, Ric. Spec. Vaticana, 8, 1
\reference Conti P., Smith, N., 2008, ApJ, 679, 1467 
\reference Crowther, P.~A. 2007, ARA\&A, 45, 177
\reference Figer, D.~F. 2005, Nature, 434, 192
\reference Gamen, R. 2004, Tesis doctoral, Univ. Nacional de La Plata
\reference van der Hucht, K.~A. 2001, New Astronomy Reviews, 45, 135
\reference Langer, N. \& Heger, A. 1999, in IAU Symposium, Vol. 193, 
	187
\reference Meynet, G. \& Maeder, A. 2005, A\&A, 429, 581
\reference Monnier, J.~D., et al. 1999, ApJL, 525, L97
\reference Niemela, V., et al. 2008, MNRAS, 389, 1447
\reference Schnurr, O., et al. 2008, MNRAS, 389, 806
\reference Shara, M., et al., 1999, \apj, 118, 390
\reference Tuthill, P.~G., et al. 1999, Nature, 398, 487

\end{referencias}

\end{document}